\documentclass[a4paper]{ESASPCS13Style}
\usepackage{epsfig, psfrag}

\newcommand \xmm {\textit{XMM-Newton}}
\newcommand \cxc {\textit{Chandra}}
\newcommand \sg {$\sigma$ Gem}

\begin{document}

\title{Flare and Quiescent X-ray Emission from Sigma Geminorum}

\author{R. Nordon\inst{1}, E. Behar\inst{1} \and M. G\"udel\inst{2}}
\institute{Department of Physics, Technion, Haifa 32000, Israel \and Paul Scherrer Institut, W\"urenlingen \& Villigen, 5232 Villigen PSI, Switzerland}

\maketitle

\begin{abstract}
The X-ray active RS CVn binary system $\sigma$ Geminorum was observed during a long-duration flare with \textit{XMM-Newton}. We compare line emission during the flare with line emission measured from a previous \textit{Chandra} observation during quiescence, in a model independent way. We find that in addition to an overall 25\% flux increase, which can be ascribed to variations in the system's quiescence activity over the 15 months that passed between the observations, there is a hot plasma component of $kT_e >$~1~keV that arises with the flare. The hot component is manifested primarily by high charge states of Fe and by a vast continuum, but with no temperature or abundance effect on the cooler ($kT_e \le$~1~keV) plasma, except perhaps for Mg that is anomalously enhanced.
\end{abstract}

\section{Introduction}
The interplay between steady coronal emission and coronal flares has been a subject of ongoing research for many years. X-ray line resolved spectra available with Chandra and XMM-Newton allows now for unprecedented plasma diagnostics, one of which is elemental abundance measurements.

The RS~CVn binary system \sg\ (HD 62044, HR 2973, HIP 37629) is X-ray bright ($ \log L_X \approx 31.0 \pm 0.2$ erg~s$^{-1}$; \cite{Yi1997}) and well observed in all wavelengths. It has a rather long period of 19.6045 days (Duemmler et al. 1997) 
for such an X-ray bright star. The primary star is a K1 III type red giant. Little is known of the secondary as it has not been detected directly at any wavelength, but restrictions to its mass and luminosity suggest that it is most likely a late type main-sequence star of one solar mass or less (\cite{Duemmler1997}).
While most previous observations found it to be a relatively steady emitter, a very large flare has been detected in December 1998 with EUVE (\cite{Sanz-Forcada2002}). Another flare was detected in April 2001, both in X-ray and in radio using XMM-Newton and the VLA (\cite{Guedel2002}).

In this poster we present a direct comparison of the soft X-ray emission from the April 2001 flare with a quiescent spectrum obtained in December 1999 in a model independent way. Our goal is to use only the line emission for the comparison as the lines constrain the plasma properties more tightly than the continuum.

\section{Observations and data analysis}

\subsection{The observations and light curves}
The target \sg\ was observed by \xmm\ with the Reflection Grating Spectrometers (RGS) on April 2001, for a total exposure time of 54 ks, of which we obtained data in the range of 6~-~38 \AA. \cxc\ observed the target on December 1999 for 100 ks with the Low Energy Transmission Grating (LETG) spectrometer in the Continuous Clocking (CC) mode for which reliable data were available only for the 2~-~20~\AA\ range. 

\begin{figure}[!t]
  \begin{center}
    \resizebox{\hsize}{!}{\includegraphics{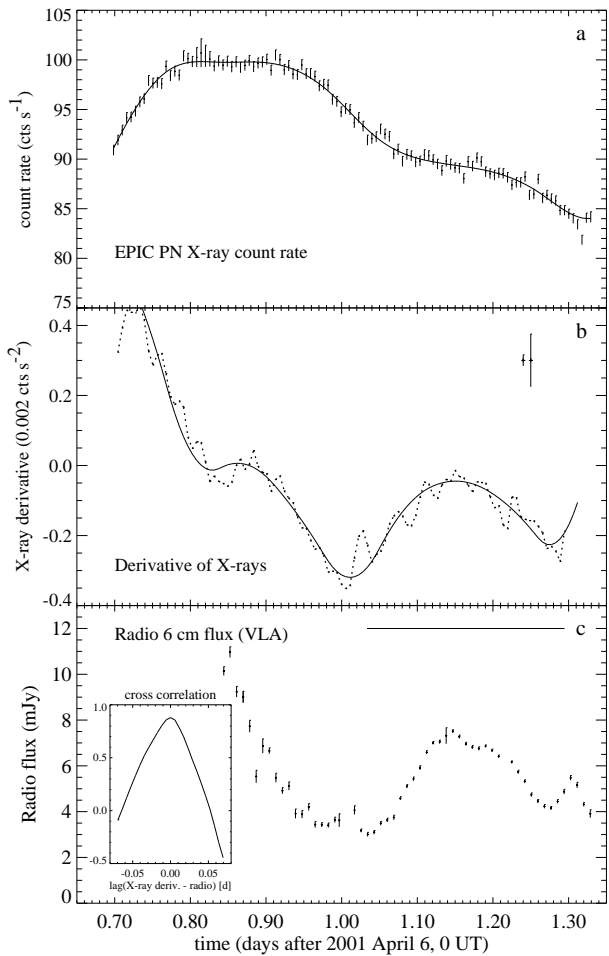}}
    \caption{Light curves of \sg, obtained on April 2001, showing the Neupert effect, reproduced from  G\"udel et al. (2002). 
(a) \xmm\ EPIC PN light curve, binned to 500~s. The smooth solid curve has been obtained from a Chebychev polynomial fit of order 9. (b) Time derivative of the smoothed X-ray light curve. Solid curve: Time derivative of the Chebychev polynomial fit above. Dotted curve: Derived from a boxcar-smoothed light curve (boxcar length = 11; not shown). For the latter derivative, the larger error bar in the upper right corner illustrates the absolute uncertainty for any single derivative, whereas the smaller error bar indicates the relative scatter between nearest neighbors and is smaller due to correlations introduced by the smoothing. (c) VLA 6~cm light curve, binned to 750~s (observing scan length). The inset shows the cross-correlation function of the radio light curve and the X-ray time derivative, computed for the time interval marked with a horizontal bar above the second radio flare.}
    \label{fg:Neupert}
  \end{center}
\end{figure}

\begin{figure}[t]
  \begin{center}
    \resizebox{\hsize}{!}{\includegraphics{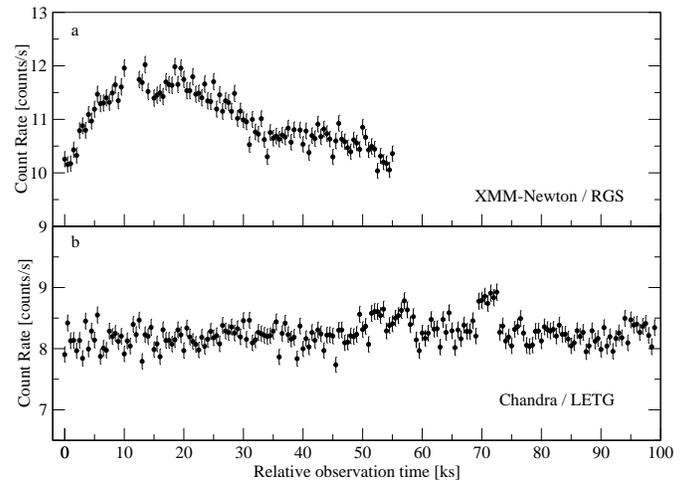}}
    \caption{Light curves for \sg\ in time bins of 500~s. (a) April 2001 RGS (1 and 2 combined) observation (1st and 2nd orders). (b) December 1999 LETG/ACIS observation (all orders).}
    \label{fg:lightcurve}
  \end{center}
\end{figure}

In the long duration flare observed with \xmm , a Neupert effect was detected using the EPIC-PN camera on board \xmm\ and the VLA. The Neupert effect supports the scenario in which the radio emission is due to fast particles spiraling down the magnetic loop, emitting gyrosyncrotron radiation, and the X-ray emission is due to chromospheric plasma heated by these particles. Thus the radio flux depends on the instantaneous number of particles spiraling down while the X-ray flux depends on the total energy transferred to the plasma, and the approximate relation is $\frac{d}{dt}L_X \propto L_{Radio}$. These results, reported by G\"udel et al. (2002), are shown in Figure~\ref{fg:Neupert}. The RGS light curve 1st and 2nd order for both instruments, rebinned to 500~s bins, is presented in Figure~\ref{fg:lightcurve}. The flare light curve can be contrasted with the flat light curve from the \cxc\ observation (using all orders of LETG), which is shown below in the same figure. Since the \cxc\ / LETG light curve is very flat we used this observation as a quiescence state reference.

\subsection{Comparison of flaring and quiescent spectra}

\begin{figure}[!ht]
  \begin{center}
    \resizebox{\hsize}{!}{\includegraphics{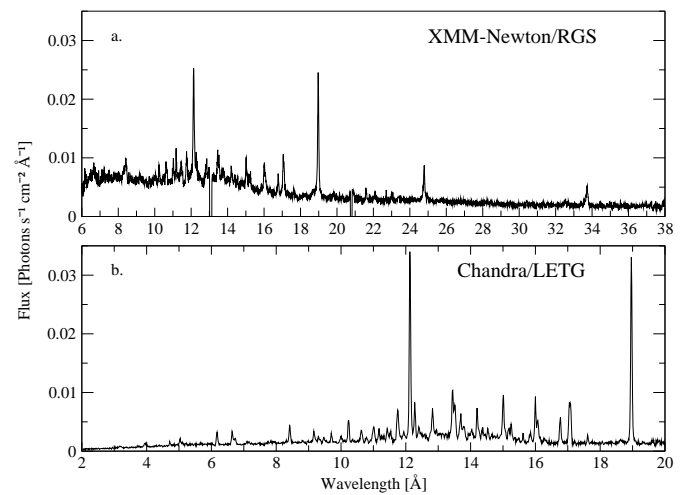}}
    \caption{\sg\ spectra. (a) Flaring \textit{XMM}-RGS spectrum, April 2001. Averaged spectrum of the two RGS instruments in 1st order. (b) Quiescent state \textit{Chandra}-LETG spectrum, December 1999. Averaged spectrum of the two 1st orders.}
    \label{fg:spectra}
  \end{center}
\end{figure}

The fluxed spectra extracted from the instruments are presented in Figure~\ref{fg:spectra}. The overlapping waveband of the instruments is only 6~-~20~\AA. The most outstanding difference is the large addition of continuum emission in the flare spectrum. The total flux in the overlapping region is increased by a factor of 2.35 in the flare spectrum, relative to the quiescent one.

In order to compare line emission, we have taken into account the different Line Spread Functions (LSF) of the instruments. LETG has approximately a Gaussian LSF with FWHM $\approx$ 0.05~\AA, while RGS has a slightly wider LSF of FWHM $\approx$ 0.07~\AA\ and a more complex shape with a narrow peak. For the comparison we fold the LETG data through the RGS response, but in order to reproduce the RGS profile correctly, we first need to represent the LETG bright lines as delta functions with their correct measured flux. This is done by Gaussian fitting to the lines with a local continuum. We then subtract the fitted Gaussians from the spectrum to get a residual continuum. Finally we fold the delta-function lines plus the residual continuum (thus ensuring the correct flux is retained with high precision even if there were some inaccuracies with the fitting) via the RGS response.

\begin{figure}[t]
  \begin{center}
    \resizebox{\hsize}{!}{\includegraphics{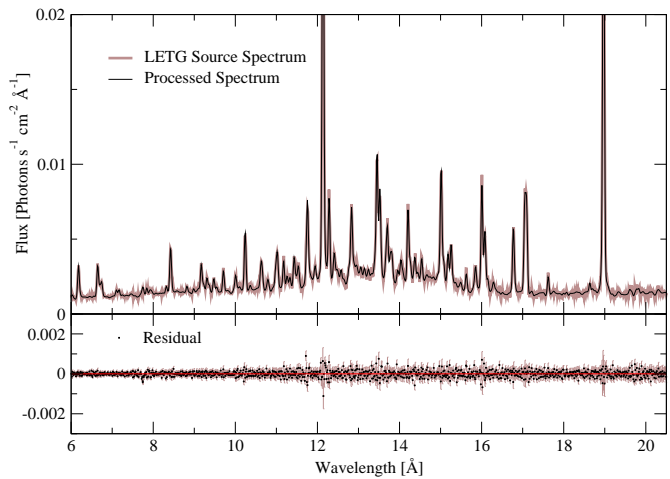}}
    \caption{The LETG source spectrum compared with the fitted Gaussians + residual continuum spectrum. The delta-functions and continuum were passed back through the LETG response and subtracted from the LETG spectrum to give the plotted residual. Random fluctuations are smoothened by to the second convolution with the LSF. The residual shows a very good match, to the order of the random fluctuations, indicating the delta-functions + continuum and the source spectrum cannot be distinguished by the LETG spectrometer (and also by RGS).}
    \label{fg:reproduced}
  \end{center}
\end{figure}

As a consistency check, we also fold the processed spectrum (delta-functions + residual continuum) back through the LETG response and compare to the original data. The results are presented in Figure~\ref{fg:reproduced} and they show that the source spectrum and the processed one cannot be distinguished by the LETG instrument. Note that in the processed spectrum, after being passed through the LETG response, the statistical fluctuations are smoothened out by the second convolution with the LSF.

In order to investigate the change in line intensities from quiescence to flare we chose to scale the emission lines by the intensity ratio of the bright, well isolated, O$^{7+}$ Ly$\alpha$ line at 19~\AA. Accordingly, the quiescent spectrum was multiplied by a factor of 1.25. To illustrate the comparison, the remaining continuum difference was also added to the quiescent spectrum to match the flaring continuum. The end result of this process is shown in Figure~\ref{fg:diff_plot} where the flare and scaled quiescent spectra are plotted one on top of the other. Note that the continuum is the flare continuum for both spectra. Only the lines should be compared. This enables a comparison of the relative line intensities directly between the two phases.

\section{Results}

\begin{figure*}[t]
  \begin{center}
    \resizebox{\hsize}{!}{\includegraphics{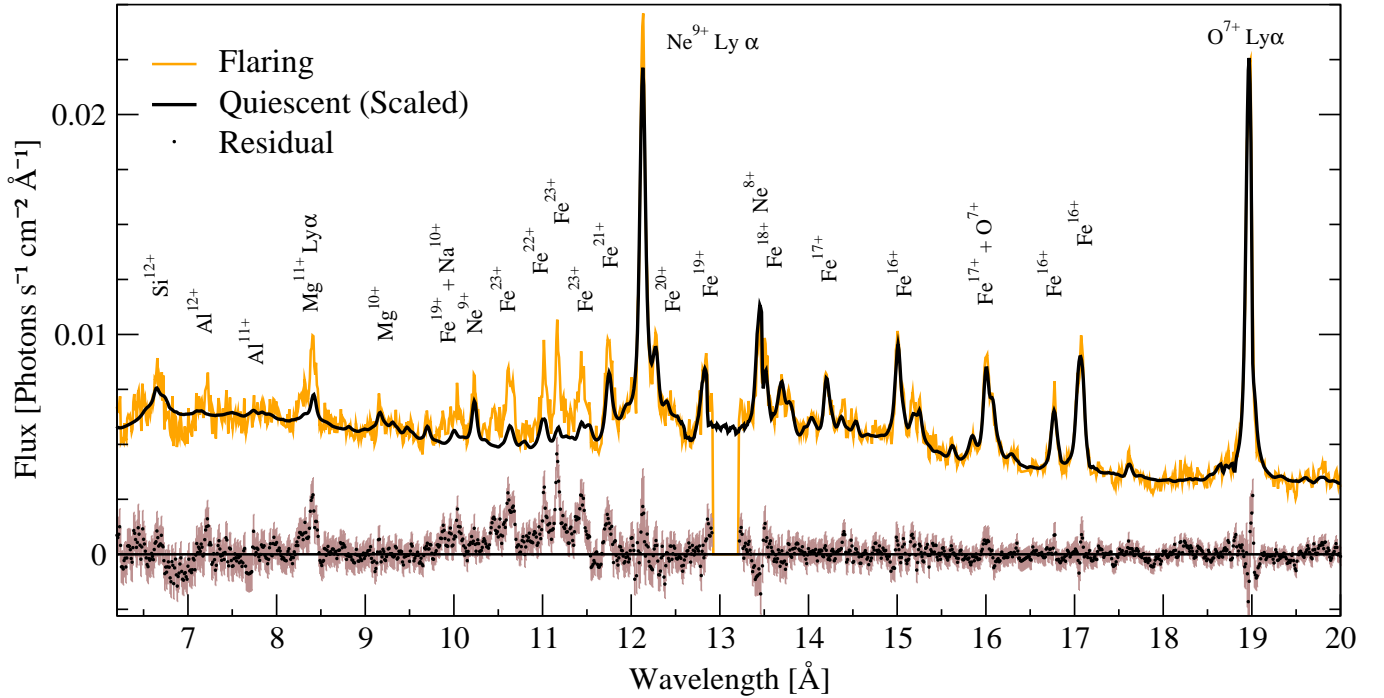}}
    \caption{Line intesities of the flaring and quiescent states compared. The quiescent lines are multiplied by a factor of 1.25 and an ad-hoc continuum is added to match the flaring continuum. The plotted residual is the quiescent curve subtracted from the flaring spectrum, with corresponding error bars.}
    \label{fg:diff_plot}
  \end{center}
\end{figure*}

As seen in Figure~\ref{fg:diff_plot}, the relative line intensities at wavelengths longer than 12~\AA\ match very well in the two spectra. The fluctuations in the residual around the brightest lines at 19, at 13.5,  and at 12~\AA\ are caused by slight inaccuracies in the LSF of the RGS intensified by the large steep peak. The residuals however average to nearly zero and in the 19~\AA\ line, exactly zero, by design of the scaling.
The lines above 12~\AA\ come from ions with maximum formation temperature of under $kT_e \sim$ 600~eV and maximum emissivity temperature (the temperature for which the emitted line is brightest) of under 650~eV. Lines at shorter wavelengths of Fe$^{19+}$ and Si$^{12+}$ at 12.84 and 6.65~\AA, respectively, whose maximum emissivity temperature is 850~eV also match well in the two spectra. 

We conclude that the ions with maximum emissivity temperatures comparable to that of Fe$^{20+}$ ($\sim$~1~keV) and below were not affected by the flare, except their emission measure increased by a constant factor of 1.25.
This means that the plasma at temperatures below $\sim$1~keV has the same abundance composition in the two observations and no change in the shape of the emission measure distribution is observed. However, the total emission measure did increase by the 1.25 factor by which we multiplied the quiescent lines. This means that either the flare is evaporating plasma from the chromosphere with the same chemical and thermal composition as the quiescent corona or, more likely, the 1.25 enhancement is unrelated to the flare and the amount of relatively cold corona increased by this factor over the 15 months that passed between the observations. Previous observations with ROSAT spanning 1.5 yrs. showed changes of the same order (\cite{Yi1997}). This might be related to the system's activity cycle or changes in active regions and not to the flare. 

More evident changes in the lines are observed for highly ionized iron lines of Fe$^{20+}$ to Fe$^{23+}$ below 12~\AA, which correspond to temperatures of 1~keV and above. Since the relative intensities of the lower-T Fe lines do not change during the flare, this is clearly a temperature effect rather than an abundance effect. This and the enhanced continuum at short wavelengths without change to the lines, indicate the appearance of a very hot component with little effect on the cooler plasma. The only other ions that probe this temperature range and appear in the RGS spectrum are Al$^{12+}$ at 7.17~\AA, which is too weak to be used and Si$^{13+}$ at 6.18~\AA, which does seem to be enhanced, but is just at the edge of the RGS spectral range and, thus, is not clear enough to be used.

Mg$^{11+}$ is the exception to the rule. Although its maximum emissivity temperature is 840~eV, it does show a large increase in the Ly$\alpha$ line at 8.42~\AA\ line, by a factor of $2.3\pm0.5$ beyond the overall 25\% enhancement. On the other hand, the other ions with similar maximum emissivity temperatures, namely Fe$^{19+}$ and Si$^{12+}$ (850~eV and 860~eV) do not show the same effect. Since Mg$^{11+}$ is a Hydrogen-like ion, its emissivity drops off slowly beyond the 840~eV peak and extends to higher temperatures than those of Fe$^{19+}$ and Si$^{12+}$. Thus, we could be seeing its emission originating at higher temperatures in the flare. The fact that no change is detected in the Mg$^{10+}$ lines which are emitted primarily at lower temperatures of $\sim$540~eV reinforces our conclusion that the cold gas is not affected by the flare.
However, since we concede that ascribing Mg$^{11+}$ to the flare is somewhat ambiguous, an abundance enhancement of Mg cannot be totally ruled out.

\section{Conclusions}
\begin{enumerate}
\item The 25\% increase in flux between 1999 and the flare of 2001 may be due to gradual variations in the soft X-ray activity rather than a result of the 2001 flare. 

\item The flare shows a two component behaviour: cooler plasma which is not effected by the flare in a detectable amount and plasma at $kT_e >$ 1~keV which is significantly enhanced. The flare is manifested unambiguously by a strong continuum that rises down to the RGS limit at 6\AA\ (while the quiescent continuum peaks around $\sim$ 13~\AA) indicating temperatures of $kT_e \ga$ 2~keV, higher than the temperatures producing the lines. This suggests that the large flare is localized and does not disturb the surrounding cooler plasma.

\item No low-T evaporation nor enrichment of the cooler parts of the corona is detected. The flare did not change the relative abundance of elements in the corona in a detectable amount, with the exception of maybe Mg.  Additional ions of Si, S, Ar, and Ca that could indicate changes at very high temperatures fall outside of the RGS band (i.e., under 6~\AA). In relation to the chromospheric evaporation scenario, this suggests that the amount of evaporated material is small relative to the quiescent corona. As the hot evaporated plasma rises from the denser chromosphere to the higher corona it expands and its emission measure drops proportionally to the electron density. Hence, the effect on the emission measure of the cooler coronal plasma is negligible and remains undetected.

\end{enumerate}

\begin{acknowledgements}
The research at the Technion was supported by ISF grant 28/03 and by a grant from the Asher Space Research Institute. PSI astronomy has been supported by the Swiss National Science Foundation (grant 20-66875.01). We thank Shai Kaspi for assistance with the data reduction. \xmm\ is an ESA science mission with instruments and contributions directly funded by ESA member states and the USA (NASA).
\end{acknowledgements}

\end{document}